\documentclass[aps, prx, twocolumn]{revtex4-2}

\usepackage{graphicx}
\usepackage{epstopdf}
\usepackage{epsfig}
\usepackage{amssymb,amsmath,stmaryrd,tabularx}
\usepackage{wrapfig}
\usepackage{bm}
\usepackage[T1]{fontenc}
\usepackage[center]{subfigure}
\usepackage[toc,page]{appendix}
\usepackage{float}
\usepackage{leqno}

\usepackage[percent]{overpic}
\usepackage{graphicx}

\newcommand{\no}{\nonumber}

\begin{document}

\title{Spectral link of the generalized Townsend-Perry constants in turbulent boundary layers}
\author{Bj\"orn Birnir $^1$,  Luiza Angheluta$^2$, John Kaminsky$^1$,  Xi Chen$^3$}
\affiliation{
 $^1$ CNLS and Department of Mathematics, University of California in Santa Barbara, USA\\
 $^2$The Njord Centre, Department of Physics, University of Oslo, P. O. Box 1048, 0316 Oslo, Norway\\
 $^3$ Institute of Fluid Mechanics, Beihang University, Beijing, China }

\begin{abstract}
We propose a minimal spectral theory for boundary layer turbulence that captures very well the profile of the mean square velocity fluctuations in the stream-wise direction, and gives a quantitative prediction of the Townsend-Perry constants. The phenomenological model is based on connecting the statistics in the streamwise direction with the energy spectrum of the streamvise velocity fluctuations. The original spectral theory was proposed in Ref. \cite{gioia2006turbulent} to explain the friction factor and von K\'arm\'an law in Ref. \cite{GGGC10}. We generalized it by including fluctuations in the wall-shear stress and the streamwise velocity.  The predicted profiles for the mean velocity and mean square fluctuations are compared with velocity data from wind tunnel experiments. 
\end{abstract}

\maketitle

\section{Introduction}

Turbulence is a ubiquitous phenomenon encountered in very diverse natural systems, from the large-scale atmosphere \cite{wyngaard1992atmospheric} and oceans \cite{toschi2009lagrangian} all the way down to quantum fluids \cite{vinen2002quantum}, as well as in engineered systems, such as pipelines, heat exchangers, wind turbines, etc. It relates to the  complex fluid dynamics that orchestrates the interactions of flow eddies spanning many length-scales and generating non-Gaussian statistics of velocity increments. The statistical properties of these turbulent fluctuations are fundamentally changed when the flow is confined by the presence of solid walls or boundaries \cite{smits2013wall,jimenez2013near}. In contrast to bulk turbulence, which is statistically homogeneous and isotropic, the wall-bounded turbulence is characterised by statistically anisotropic properties. Namely, there is a net mean-flow in the streamwise direction along the wall and the different flow structures form depending on their distance to the wall. We typically differentiate between four flow regions as moving away from the wall \cite{Ob97}: i) the {\em viscous region} closest to the wall and where viscous flows dominate, ii) the {\em buffer layer}, marking the transition from the viscous layer into the inertial layer, iii) the {\em inertial layer} where the 
log-law of the wall applies, and iv) the {\em wake}, the energetic region beyond the inertial layer. A more refined division is given in \cite{CHS19}. 

A classical signature of wall-bounded turbulence is the "log-law of the wall" of the mean velocity profile (MVP) due to Prandtl and von K\'arm\'an, and reads as 
\begin{equation}
\label{eq:l-lwall}
\langle \tilde u \rangle = \frac{1}{\kappa} \log(\tilde y)+B,
\end{equation}
where $\kappa$ is the \emph{universal} von K\'arm\'an constant that is independent of the microscopic flow characteristics and relates to generic features such as space dimensionality.  The distance to the wall $y$ and the mean fluid velocity $u$  along the wall, 
are typically expressed in  the "wall units" determined by the wall shear stress $\tau_0$. This is because $\tau_0$ is an important theoretical concept that is also experimentally measurable. The friction velocity $u_\tau = \sqrt{\langle \tau_0 \rangle/\rho}$ which is set by the wall shear stress $\tau_0$ and the kinematic viscosity $\nu$, and enters in the unit rescalings as $\tilde u= u/u_\tau$ and $\tilde y = yu_\tau/ \nu$. The constant fluid density is $\rho$ and the $B$ is a dimensionless constant that
is fitted to experimental data, e.g.  \cite{P53}. 

\begin{figure}[t]
\centering
  \includegraphics[width=.4\textwidth]{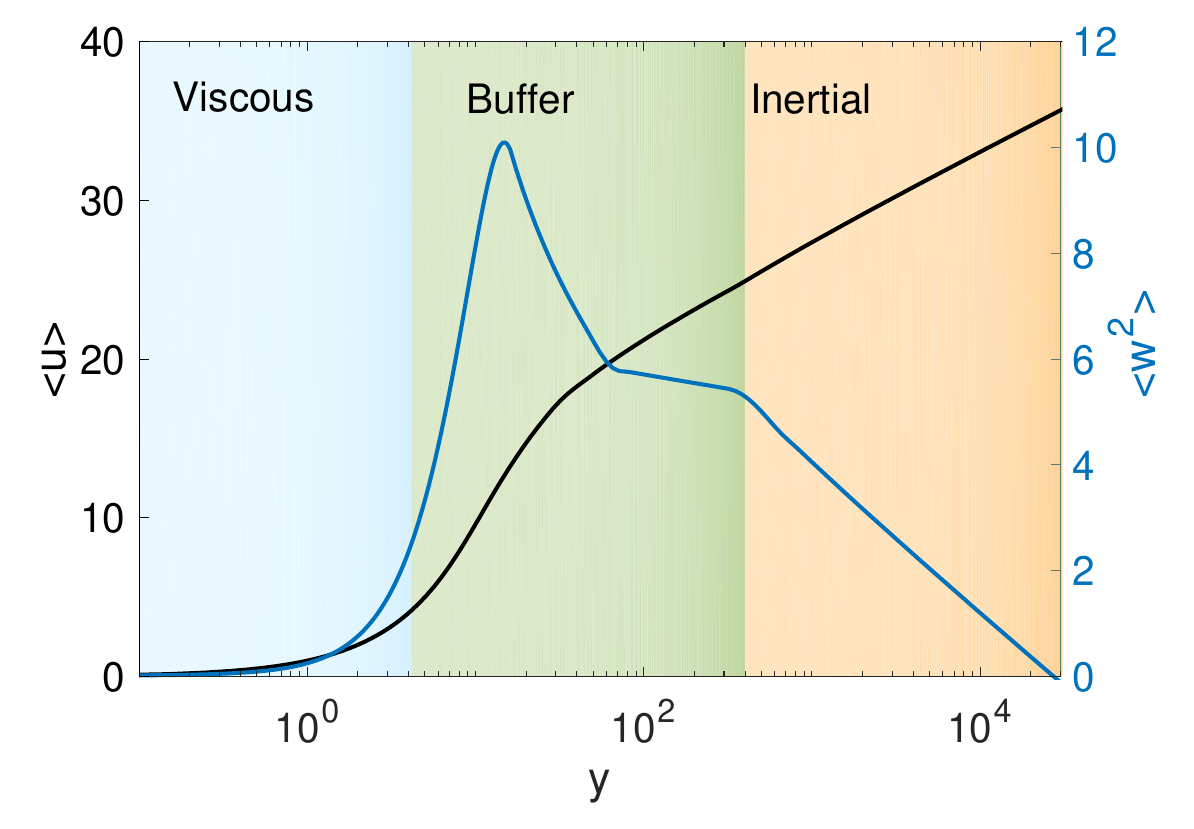}
  \caption{Theoretical predictions from the spectral theory for the MVP $\langle u \rangle$  and mean square velocity fluctuations $\langle w^2\rangle$ (dimensionless variables in wall units). }
  \label{fig:wx1}
\end{figure}

A log-law of the wall was also derived from the "attached eddy hypothesis" by Townsend \cite{T76}.  Townsend showed that the velocity fluctuations, $\tilde w =w/u_\tau$,  $\tilde u= \langle \tilde u \rangle + \tilde w$, also follow the log-law of the wall in its second moment, namely
\begin{equation}
\label{eq:l-lfluct}
{\langle \tilde w^2 \rangle} = - A_1 \log(\tilde y) + B_1, 
\end{equation}
where the coefficients $A_1$ and $B_1$, also called the Townsend-Perry constants, were first measured by Perry and Chong \cite{PC82, P86}. 

More recently, the log-law was generalised to any moment of the streamwise velocity fluctuations, $\tilde w$, assuming Gaussian velocity fluctuations \cite{MM13},
\begin{equation}
\label{eq:l-pfluct}
\langle \tilde w^{2p} \rangle^{1/p} = - A_p \log(\tilde y) + B_p.
\end{equation}
While the generalised log-law is supported by wall-turbulence experiments, the dependance of $A_p$ and $B_p$ on $p$ turns out to be sub-Gaussian, which is confirmed both experimentally and numerically, \cite{MM13}. The sub-Gaussian behavior was explained in Ref. \cite{BC16} using the stochastic closure theory of turbulence \cite{BB211,BB314} and the analysis was improved in Ref. \cite{KBK19}, using measurements from the Flow Physics Facility (FPF) at the University of New Hampshire. 
Both of these studies used the results from homogeneous turbulence \cite{KBBS17} and made an assumption about the form of the fluctuating shear stress in the inertial layer, based on physical principles.

In  Ref. \cite{GGGC10}, a spectral theory for the log-law of the wall of the MVP was proposed in which it is possible to derive the log-law in the inertial layer and the laminar profile in the viscous layer. The novel contribution is the precise form of the transition in the buffer layer using the the Kolmogorov-Obukhov energy spectrum of turbulent fluctuations. The form of the MVP in the wake is also obtained. This was done by summing the energy of the wall-attached eddies, as hypothesised originally by Townsend in \cite{T76}.

In this paper, we propose a generalisation of the spectral theory  that includes fluctuations in the streamwise velocity due to an essentially fluctuating wall shear stress. Fig.~\ref{fig:wx1} shows the spectral theory predictions of the profiles of the mean velocity and mean square velocity fluctuations across the viscous, buffer and inertial layers. 
 The rest of the paper is structured as follow.  We summarise the analysis in Ref. \cite{GGGC10} and its extension in Section \ref{sec:MVP}, and generalise it to include the fluctuations in Section \ref{sec:fluct}. This produces the log law of the wall in  Eq. (\ref{eq:l-lfluct}) for the velocity fluctuations and its higher moments in Eq. (\ref{eq:l-pfluct}). Then in Section \ref{sec:functional}, we derive the functional form of the mean-square fluctuations in the viscous layer and the inertial layer. In Section \ref{sec:SCT}, we use the attached eddy hypothesis and the stochastic closure theory \cite{BB211,BB314} to derive the form of the Townsend-Perry and the generalized Townsend-Perry constants. This allows us to derive the streamwise fluctuations in the wall shear stress, and remove the assumption made in Refs. \cite{BC16} and \cite{KBK19}, and mentioned above. Using theory-informed by data analysis, we can construct the Townsend-Perry constants and the generalised Townsend-Perry constants. In Section \ref{sec:BW}, we extend the formulas for the mean square fluctuations to the buffer layer and the energetic wake.  In Section \ref{sec:data}, we compare the predicted MVP and mean-square velocity profile from this spectral theory to experimental data. In Section \ref{sec:summary}, we conclude with a discussion on the proposed spectral theory and the role that Townsend's attached eddies play in it.

\section{The Spectral Theory}
\label{sec:MVP}
The typical velocity of an inertial eddy of size  $s$ can be obtained by integrating out the kinetic energy contained in all eddies of sizes up to $s$ as in Ref. \cite{GGGC10} 
\begin{equation}\label{eq:vs}
v_s^2 = \int_{1/s}^\infty E(k)dk,
\end{equation}
where kinetic energy spectrum follows the Kolmogorov-Obukhov scaling with cutoffs in the injection scale and viscous scales, $E(k) =c_d(\eta k) \frac{2}{3}(\kappa_\epsilon \epsilon )^{2/3}k^{-5/3} c_e(Rk)$, with $\frac{2}{3}(\kappa_\epsilon \epsilon)^{2/3}k^{-5/3}$ being the Kolmogorov-Obukhov spectrum and $c_d (\eta k)$ and $c_e(R k)$  the phenomenological dimensionless corrections functions in the dissipative (set by the Kolmogorov scale $\eta$) and energetic range (set by the system size $R$), respectively. $\kappa_\epsilon$ is a dimensionless parameter, $\epsilon$ is the turbulent energy dissipation rate, $\eta=\nu^{3/4}\epsilon^{-1/4}$ is the viscous length scale and $R$ is the largest length scale in the flow.  
The dissipative correction function is typically an exponential cutoff function $c_d(\eta k) =\exp(-\beta_d \eta k)$, and the energetic-range
(wake) correction function is $c_e(Rk)=(1+(\beta_e/(Rk))^2)^{-17/4}$, which is the form that was proposed by von K\'arm\'an. $\beta_d$ and $\beta_e$ are non-negative fitting parameters that can be adjusted to data.
By the change of variables $\xi = sk$, we recast Eq. (\ref{eq:vs}) as
\begin{equation}
v_s^2 = (\kappa_\epsilon \epsilon s)^{2/3}I\left(\frac{\eta}{s},\frac{s}{R}\right),
\end{equation}
where the spectral function $I$ is given by the formula \cite{GGGC10}
\begin{align}
\label{eq:spectcont}
&I\left(\frac{\eta}{s},\frac{s}{R}\right)=  \nonumber \\
&\frac{2}{3} \int_1^\infty e^{-\xi \beta_d \eta/s}\xi^{-5/3}\left(1+\left(\frac{\beta_e s}{R\xi}\right)^2\right)^{-17/6} d\xi.
\end{align}
The integral sums the energies of all eddies of a smaller radius than $s$, and computes their contribution to the energy of the eddy of radius $s$. This is the energy (or spectral) formulation of the attached eddy hypothesis of Townsend \cite{T76}. The $I$-function correctly captures the buffer layer, as the transition from the viscous to the inertial layer, and the asymptotic of the MVP in the energetic wake. The asymptotic values are such that in the inertial layer $I=1$ and in the viscous layer $I=0$. The $I$-function combines the Kolmogorov-Obukhov theory with the observed spectrum in the viscous layer, the inertial layer and the wake and is thus able to capture the transition from one layer to the next. In Ref \cite{GGGC10}, it was used to give the details of the MVP. In this paper, we will use it to capture the profile of mean-square fluctuations.  

In the buffer layer a different scaling of the attached eddies comes into play, this is the $k_x^{-1}$ scaling of the spectrum that has been debated in literature, but clearly shows up in recent simulations and experiments in the middle of the buffer layer, see Figure 9 (a) in Ref. \cite{LM15} and Figure 12 (b) in Ref. \cite{Sa18}. In the spectral theory,  corresponding $I$-function for this scaling regime is
\begin{align}  
\label{eq:spectcont1}
&I_b\left(\frac{\eta}{s},\frac{s}{R}\right)= \nonumber\\
&\frac{2}{3}s^{-\frac{2}{3}} \int_1^\infty e^{-\xi \beta_d \frac{\eta}{s}}\xi^{-1}\left(1+\left(\frac{\beta_e s}{R\xi}\right)^2\right)^{-\frac{17}{6}} d\xi, 
\end{align}
where the subscript $b$ stands for "buffer". The mean velocity is primarily influenced by the $I$-function, whereas the variation (fluctuation squared) is greatly influenced by the $I_b$-function in the buffer layer. $I$ is associated with the Kolmogorov-Obukhov energy cascade $k_x^{-5/3}$, in the inertial layer, whereas $I_b$ is associated with the $k_x^{-1}$ scaling in the buffer layer. (Here the $x$ denotes the streamwise direction.) We will take $I_b$ to be zero outside the buffer layer. 

The splitting of the near-wall region based on different scaling of the spectrum was proposed by Perry and Chong \cite{PC82} who used it build an interpolation model for MVP and the variation, this model was improved in Ref. \cite{Va15}. 

\section{The generalised log-law}
\label{sec:fluct}
In this section, we will give a simple derivation of the log-law for the 
mean-square velocity profile that holds in the limit of large Reynolds number. 
In the following section we derive the general form of the variation that is not equally transparent. 

We will generalize the derivation of the MVP in Ref. \cite{GGGC10}, by adding a fluctuation to the mean velocity. We let the velocity along the wall be
\begin{equation}
v_1=u+v_1-u=u+w,
\end{equation}
where $u$ is the mean velocity obtained by  averaging $v_1$ over time, and $w$ is the fluctuation.
The same derivations as in Ref. \cite{GGGC10} give the following equations for a dominant eddy of radius 
$s=y$, if we include the velocity fluctuations. In Ref. \cite{GGGC10} the shear stress at the distance $y$ from the wall is given
by the formula ${\bar \tau_t} = \kappa_\tau \rho y v_y u'$ where $u'$ denotes the $y$ derivative of the velocity $u$ along the wall, and the overline indicates a not-fluctuating quantity. When velocity fluctuations are included the shear stress becomes:
\begin{equation}
\label{eq:shear-stress}
\tau_t = \kappa_\tau \rho y v_y (u'+w'),
\end{equation}
where $\rho$ is the density $v_y$ is the (rotational) velocity of an eddy  a distance $y$ from the wall
and $\kappa_\tau$ is the dimensionless proportionality factor.
The energy dissipation rate is related to the wall shear stress as ${\bar \epsilon} = \tau_t u'/\rho$  \cite{GGGC10} , and including the fluctuations, this becomes 
\begin{equation}
\label{eq:energy}
\epsilon = \tau_t(u'+w')/\rho.
\end{equation}
The eddy velocity for an eddy with radius $s=y$ at the distance $y$ from the wall is the same 
as in Ref. \cite{GGGC10}, and as discussed above,
\begin{equation}
\label{eq:e-viscosity}
v_y= (\kappa_\epsilon \epsilon y)^{1/3} \sqrt{I},
\end{equation}
where $I$ is the integral from Eq. (\ref{eq:spectcont}) and $\kappa_\epsilon$ is a dimensionless proportionality factor.
In the inertial layer $I=1$ and $\kappa_\epsilon = 4/5$ according to Kolmogorov's $4/5$ law. 

 Eliminating $\epsilon$ and $v_y$ from the three equations above, we obtain 
\begin{equation}
\label{eq:shear-stress_1}
\tau_t= (\kappa_\epsilon \kappa_\tau^3)^{1/2} \rho y^2 (u'+w')^2 I^{3/4}.
\end{equation}
The viscous shear stress is $\rho \nu (u'+w')$ so the total shear stress, including the contribution from the 
fluctuation is \cite{T76}
\begin{equation}
\tau_t + \rho \nu (u'+w') =  \tau_0(1-y/R).
\end{equation}
Our assumption is that the wall shear stress $\tau_0$ is also a quantity that fluctuates about its mean value. 

We change the rescaled variables in the wall units written here in terms of the friction factor $f$: $\tilde y=y Re\sqrt{f}/R$, $\tilde u = u/(U\sqrt{f})$ and $\tilde w=w/(U\sqrt{f})$ and let $f=\langle \tau_0\rangle/\rho U^2$. 
Then, the equation above becomes 
\begin{equation}
\label{eq:total_stress1}
{\tilde \kappa}^2 {\tilde y}^2(\tilde u'+\tilde w')^2 I^{3/4}+(\tilde u'+\tilde w') = \frac{\tau_0}{\langle \tau_0 \rangle}\left(1-\frac{\tilde y}{Re\sqrt{f}}\right).
\end{equation}
If we let $\tilde y \to 0$, $\tilde w \to 0$ and integrate, we get the law of the viscous layer
\begin{equation}
\label{eq:viscous}
\tilde u = \tilde y, 
\end{equation}
the laminar profile being
\begin{equation}
\label{eq:laminar}
\tilde u = \left(\tilde y-\frac{\tilde y^2}{2Re\sqrt{f}}\right). 
\end{equation}
In the large Reynolds number limit, solving just for the mean velocity, we obtain the Prandtl-von K\'arm\'an law 
\begin{equation}
\label{eq:velocity}
\tilde u = \frac{1}{\tilde \kappa}\log (\tilde y)+D.
\end{equation}
This is the correct leading term but the full formulas in the next section are more complicated. 
We now motivate the log-law for the variation. 
If we solve for both the mean velocity and the fluctuation in the large Reynolds number limit, 
we get that
\begin{equation}
\label{eq:velocity1}
\tilde u+\tilde  w= \frac{\sqrt{\tau_0}}{\langle \tau_0\rangle^{1/2} \tilde \kappa} \log (\tilde y)+C.
\end{equation}
This is consistent with the Eq. (\ref{eq:velocity}) in the sense that if $\sqrt{\tau_0}=\langle \tau_0\rangle^{1/2}$, then
$\tilde w =0$ and we recover Eq. (\ref{eq:velocity}).
Thus squaring Eq. (\ref{eq:velocity1}) gives that
\begin{equation}
{\tilde u}^2+2\tilde u \tilde w +{\tilde w}^2= \frac{\tau_0}{\langle \tau_0\rangle \tilde \kappa^2} (\log(\hat y))^2+2\frac{\sqrt{\tau_0}}{\tilde \kappa \sqrt{\langle \tau_0 \rangle}} C\log(\tilde y) +C^2.
\end{equation}
Taking the average, using that $\langle \tilde w \rangle = 0$ and Eq. (\ref{eq:velocity}), we get that
\begin{equation}
\langle \tilde w^2 \rangle = \frac{2C\langle \sqrt{\tau_0}\rangle-2D\sqrt{\langle\tau_0\rangle}}{\tilde \kappa \sqrt{\langle \tau_0 \rangle}}\log(\tilde y) +C^2-D^2.
\end{equation}
By comparing this with the generalised log-law in Eq. (\ref{eq:l-lfluct}), for the fluctuations squared, we obtain
\begin{equation}
\label{eq:gloglaw}
\langle \tilde w^2 \rangle = -A \log(\tilde y)+B,
\end{equation}
where $A = - \frac{2C\langle \sqrt{\tau_0}\rangle-2D\sqrt{\langle\tau_0\rangle}}{\tilde \kappa \sqrt{\langle \tau_0 \rangle}}$ and $B = C^2-D^2$ are the Townsend-Perry constants. The full formulas in next section show that Eq. (\ref{eq:gloglaw}) is the leading term and  $A = - 2C(\frac{\langle \sqrt{\tau_0}\rangle-\sqrt{\langle\tau_0\rangle}}{\tilde \kappa \sqrt{\langle \tau_0 \rangle}})$, 
with $C=D$. 

To simplify the notation, we will now drop the tilde's from all the variable with the dimensionless units implicitly assumed, unless otherwise stated.  

\section{The functional form of the Townsend-Perry law}
\label{sec:functional}

We will now use Eq. (\ref{eq:total_stress1}) to find the general form of the average of the fluctuations squared as a function of the distance to the wall. We consider the Eq. (\ref{eq:total_stress1})
\begin{equation}
\label{eq:total_stress2}
{\kappa}^2 {y}^2( u'+ w')^2 I^{3/4}+(u'+w') = \frac{\tau_0}{\langle \tau_0 \rangle}(1-\frac{y}{Re\sqrt{f}}),
\end{equation}
and first set  $I=0$ in the viscous layer. Then
\begin{equation}
\label{eq:u_o}
u = y - \frac{y^2}{2 Re\sqrt{f}}
\end{equation}
by averaging and integration in $y$. Integrating  Eq. (\ref{eq:total_stress2}) and subtracting $u$ gives,
\begin{equation}
\label{eq:w_o}
w = \frac{\tau_0 - \langle \tau_0 \rangle}{\langle \tau_0 \rangle}\left( y - \frac{y^2}{2 Re\sqrt{f}}\right)
\end{equation}
and
\begin{equation}
\langle w^2 \rangle=\frac{\langle \tau_0^2 \rangle - \langle \tau_0 \rangle^2}{\langle \tau_0 \rangle^2}\left( y - \frac{y^2}{2 Re\sqrt{f}}\right)^2.
\end{equation}

In the inertial layer $I=1$ and ignoring the small $O(1/y^4)$ term, we get that 
\begin{align}
u+w &=& \frac{1}{2\kappa^2 y} + 2\frac{\sqrt{\tau_0}}{\kappa \sqrt{\langle \tau_0 \rangle}}\sqrt{1-\frac{y}{2Re\sqrt{f}}} \nonumber\\
&-&2 \frac{\sqrt{\tau_0}}{\kappa \sqrt{\langle \tau_0 \rangle}} \tanh^{-1}\left(\sqrt{1-\frac{y}{2Re\sqrt{f}}}\right)+K, 
\end{align}
where $K$ is a constant. Then setting $w=0$, we get that 
\begin{align}
&&u = \frac{1}{2\kappa^2 y} + \frac{2}{\kappa}\sqrt{1-\frac{y}{2Re\sqrt{f}}}\nonumber\\
&-&\frac{2}{\kappa}\tanh^{-1}\left(\sqrt{1-\frac{y}{2Re\sqrt{f}}}\right)+K',
\end{align}
where $K'$ is another constant, because $\tau_0$ becomes $\langle \tau_0 \rangle$. Subtracting, $u$ from $u+w$ we get
\begin{align}
&&w =  2\frac{(\sqrt{\tau_0}-\sqrt{\langle \tau_0 \rangle})}{\kappa \sqrt{\langle \tau_0 \rangle}}\sqrt{1-\frac{y}{2Re\sqrt{f}}}\no\\
&-&2 \frac{(\sqrt{\tau_0}-\sqrt{\langle \tau_0 \rangle})}{\kappa \sqrt{\langle \tau_0 \rangle}} \tanh^{-1}\left(\sqrt{1-\frac{y}{2Re\sqrt{f}}}\right)+C,
\end{align}
where $C=K-K'$. Squaring $w$ and taking the average gives
\begin{align}
&&\langle w^2 \rangle =  4C\frac{(\langle \sqrt{\tau_0} \rangle-\sqrt{\langle \tau_0 \rangle})}{\kappa \sqrt{\langle \tau_0 \rangle}}\sqrt{1-\frac{y}{2Re\sqrt{f}}} \no\\
&-&4C \frac{(\langle \sqrt{\tau_0} \rangle-\sqrt{\langle \tau_0 \rangle})}{\kappa \sqrt{\langle \tau_0 \rangle}} \tanh^{-1}\left(\sqrt{1-\frac{y}{2Re\sqrt{f}}}\right)\no\\
&+&4\left[\frac{2(\langle\tau_0 \rangle-\sqrt{\langle \tau_0 \rangle}\langle \sqrt{\tau_0}\rangle)}{\kappa^2 \langle \tau_0 \rangle}\left(1-\frac{y}{2Re\sqrt{f}}\right.\right.\no\\ 
&-& \left. 2\sqrt{1-\frac{y}{2Re\sqrt{f}}}\tanh^{-1}(\sqrt{1-\frac{y}{2Re\sqrt{f}}})\right)\no\\
&+& \left.  \left[\tanh^{-1}(\sqrt{1-\frac{y}{2Re\sqrt{f}}})\right]^2\right]+C^2.
\end{align}
From $\tanh^{-1}(x) = \frac{1}{2} \log(\frac{1+x}{1-x})$, we see that the second term in the last formula is of leading order and we get that
\begin{equation}
\label{eq:w2der}
\langle w^2 \rangle \sim 2C \frac{(\langle \sqrt{\tau_0} \rangle-\sqrt{\langle \tau_0 \rangle})}{\kappa \sqrt{\langle \tau_0 \rangle}}\log\left(\frac{y}{Re\sqrt{f}}\right) + h. o. t. 
\end{equation}
This agrees with the formula (\ref{eq:gloglaw}) above. For higher order moments $\langle w^{2p} \rangle^{1/p}$ the similar term, 
linear in $\tanh^{-1}$ and multiplied by $2C$, is  of leading order,
\begin{equation}
\label{eq:wpder}
\langle w^{2p} \rangle^{1/p} \sim 2C \frac{\langle (\sqrt{\tau_0}-\sqrt{\langle \tau_0 \rangle})^p \rangle^{1/p}}{\kappa \sqrt{\langle \tau_0 \rangle}}\log\left(\frac{y}{Re\sqrt{f}}\right) + h. o. t. 
\end{equation}
These formulas establish the log dependance of the second moment of the fluctuations, with the Townsend-Perry constants, and the log dependence of the higher moments of the fluctuations, with the Generalized Townsend-Perry constants,
and justify formulas Eq. (\ref{eq:l-lfluct}) and Eq. (\ref{eq:l-pfluct}). Together, Eq. (\ref{eq:l-lfluct}) and Eq. (\ref{eq:l-pfluct}) can be called the generalised log-law of the wall. 

\section{Derivation of the Generalized Townsend-Perry Constants}
\label{sec:SCT} 
\begin{figure}[h]
\centering
  \includegraphics[width=.3\textwidth]{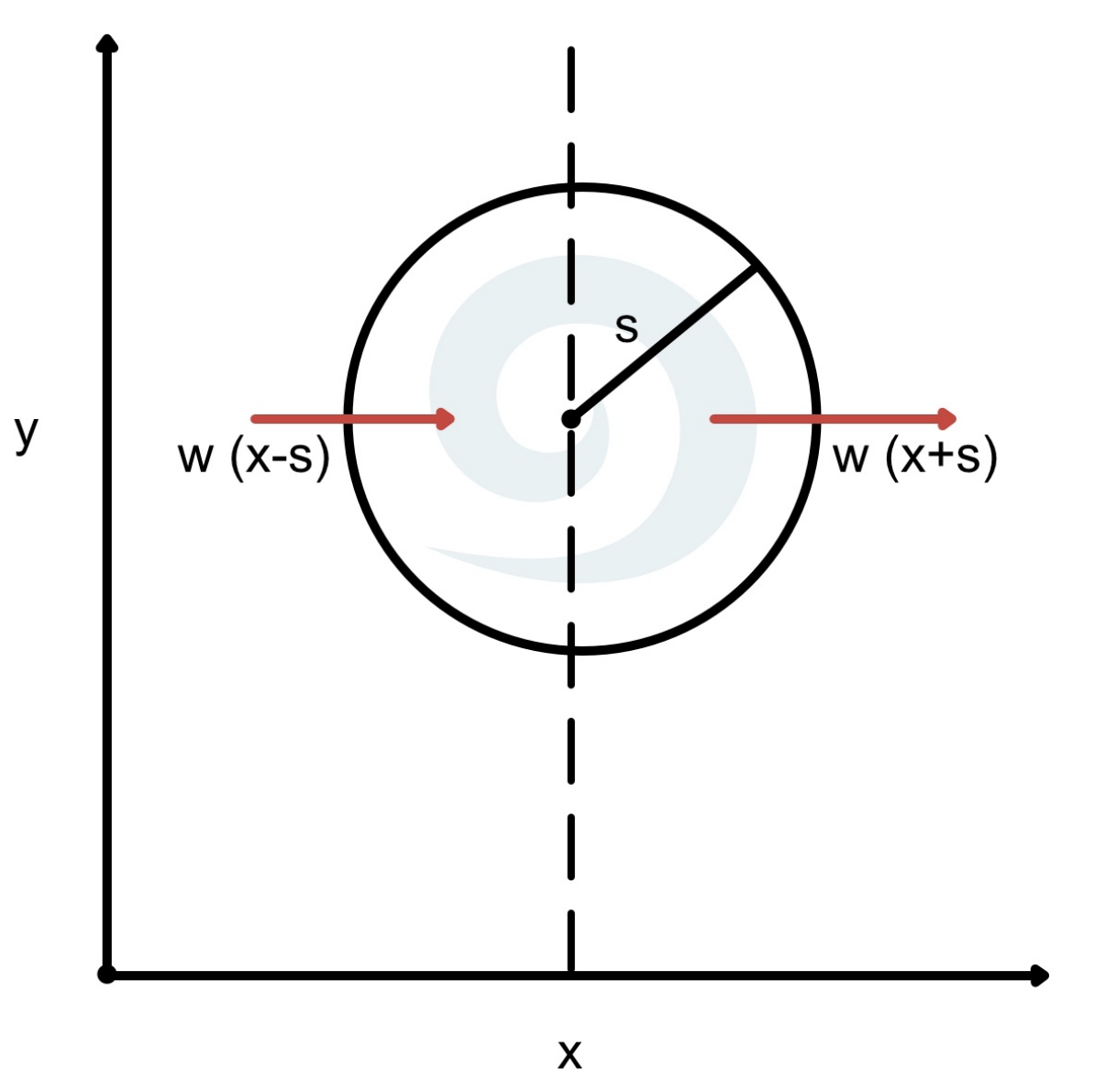}
  \caption{The eddy of radius $s$ and the variation in the fluctuations across it in the $x$ (streamwise) direction.}
  \label{fig:wx}
\end{figure}

We consider the dependence of the fluctuation $w$ on the distance $x$ along the wall, to understand the Townsend-Perry constants. So far we have only considered $w(y)$ as a function of the distance $y$ from the wall, but $w(x,y)$ obviously depends on both variables $x$ and $y$. If we consider the eddy depicted in Fig. \ref{fig:wx}, then we see that the difference in momentum in the $x$ direction, across the eddy, is given by 
\begin{equation}
\rho(w(x+s)-w(x-s)) \sim 2\rho s w_x,
\end{equation}
for $y$ fixed, where $w_x=\frac{d}{dx}w$. 

This means that the total turbulent stress, across a vertical surface at $x$, denoted by a dotted line on Fig. \ref{fig:wx} for an eddy of radius $s\sim y$, is 
\begin{equation}
\tau_0 = \tau_t+\tau_x,
\end{equation}
where $\tau_x= 2\kappa_\tau \rho y w_x v_y$, analogous to formula Eq. (\ref{eq:shear-stress}) above. Then  we get, using 
Eq. (\ref{eq:e-viscosity}) and
\begin{equation}
\epsilon =  (\tau_t+\tau_x)(u'+w_x) \rho,
\end{equation}
that
\begin{equation}
\tau_t+\tau_x= \kappa^2 \rho I^{3/4} y^2(u'+w_x)^2,
\end{equation}
where prime denotes the derivative with respect to $y$, and
\begin{align}
(\tau_t+\tau_x)^{1/2} &=& \kappa \rho^{1/2}I^{3/8}y(u'+w_x) \no\\
&=& \langle \tau_0 \rangle^{1/2}+ \kappa \rho^{1/2}I^{3/8}y|w_x|,
\end{align}
since both parts must be positive. The derivation is completely analogous to the derivation in Sec. \ref{sec:fluct}, 
but here with $w$ varying in the $x$ direction and $w_y=0$. 
This gives that for $y$ fixed,
\begin{align}
\tau_0^{1/2}-\langle \tau_0 \rangle^{1/2} &=& (\tau_t+\tau_x)^{1/2}-\langle \tau_0 \rangle^{1/2} \no\\
&=& \kappa \rho^{1/2}I^{3/8}y|w_x|.
\end{align}
Considering the leading order $\log(y/2Re\sqrt{f})$ term in Eq. (\ref{eq:w2der}) gives the Townsend-Perry constant
\begin{equation}
\label{eq:TP}
A_1=\frac{2C \rho^{1/2} y\langle |w_x|\rangle}{\sqrt{\langle \tau_0 \rangle}},
\end{equation}
and the generalized Townsend-Perry constants
\begin{equation}
\label{eq:GTP}
A_p=\frac{2C \rho^{1/2} y \langle |w_x|^{p}\rangle^{1/p}}{\sqrt{\langle \tau_0 \rangle}},
\end{equation}
by use of  Eq. (\ref{eq:wpder}). This justifies the form of the stress tensor assumed in Ref. \cite{BC16} and used in  Ref. \cite{KBK19}.
 Finally, we get the expressions
\begin{equation}
A_1 = K \langle |w(x+y)-w(x-y)|\rangle
\end{equation}
and
\begin{equation}
A_p = K \langle |w(x+y)-w(x-y)|^{p} \rangle^{1/p},
\end{equation}
where $K$ is a constant and this produces the relationship between the Townsend-Perry and the generalized Townsend-Perry constants and the structure function of turbulence, see Ref. \cite{BB211,BB314,KBBS17}, used in Ref. \cite{BC16,KBK19},
\begin{equation}
\label{eq:TPstru1}
A_1 = K C_1|y^*|^{\zeta_1},
\end{equation}
\begin{equation}
\label{eq:TPstru2}
A_2 = K C^{1/2}_2|y^*|^{\zeta_2/2},
\end{equation}
and 
\begin{equation}
\label{eq:TPstrup}
A_p = K C^{1/p}_{p}|y^*|^{\zeta_{p}/p},
\end{equation}
where $-y\leq y^* \leq y$. 
Considering the ratio, washes out the constant $K$, 
\begin{equation}
\label{eq:ratio}
\frac{A_p}{A_2}= \frac{C^{1/p}_{p}}{C^{1/2}_2}|y^*|^{\zeta_{p}/p-\zeta_2/2},
\end{equation}
where the $C_p$s are the Kolmogorov-Obukhov coefficients of the structure functions from Ref. \cite{BB211,BB314,KBBS17}.
The last ratio was used in Ref. \cite{KBK19} to get agreement between experimental data and theory.

\begin{figure}[t]
\centering
  \includegraphics[width=0.45\textwidth]{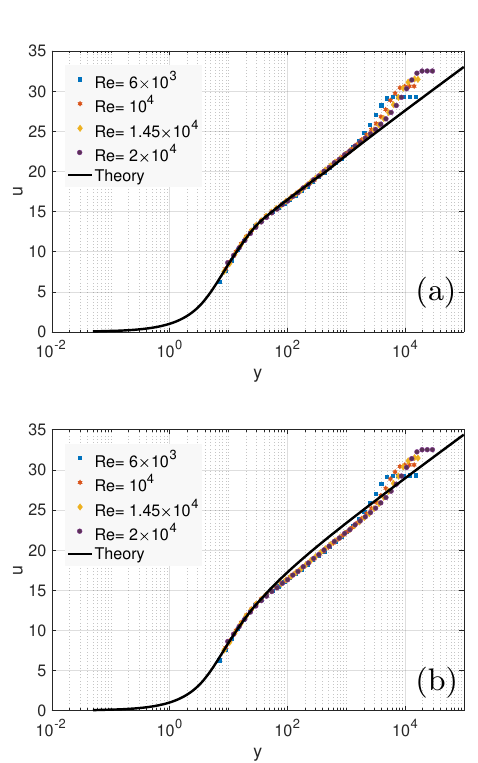}
  \caption{The average of the MVP as a function of $log(y)$, where $y$ is the distance from the wall. Comparison of experimental data with theory (black line). (a) Theoretical curve is given by an $I$-integral that interpolates between the $k_x^{-5/3}$ to the $k_x^{-1}$ with $a=0.9994$ in the buffer region. (b) Theoretical curve has a uniform $I$-integral with the $k_x^{-5/3}$ scaling present in buffer and inertial regions.}
  \label{fig:mean velocity}
\end{figure}

\section{The Spectral Theory of mean-square fluctuations}
\label{sec:BW}
In the above sections we have not used the spectral information in the integral $I$, in Eq. (\ref{eq:spectcont}). We have just used the attached eddy hypothesis and set $I=0$ in the viscous layer and $I=1$ in the inertial layer. But following 
Ref. \cite{GGGC10}, we can now use the spectral information through the integral $I$ to find the beginning of the buffer layer and the form of both the MVP $u$ and the fluctuation $w$ in the buffer layer and in the wake. This allows one to obtain the full functional form of both $u$ and $w$ as functions of the distance $y$ from the wall and compare it with the experimental data in the next section. By use of the energy Eq. (\ref{eq:energy}) and the relation $\eta = \nu^{3/4}\epsilon^{-1/4}$ we can find an expression for $\eta/y$, the viscosity parameter that increases as we approach the wall $y\to 0$. If we set the fluctuation equal to zero, 
\begin{equation}
\eta/y = (\tilde u'(1-\tilde y/Re\sqrt{f})-(\tilde u')^2)^{-1/4}\tilde y^{-1}
\end{equation}
and find a formula for $\tilde y$ using this equation along with the equation
\begin{equation}
{\kappa}^2 {\tilde y}^2( u')^2 I^{3/4}+u' = \frac{\tau_0}{\langle \tau_0 \rangle}\left(1-\frac{y}{Re\sqrt{f}}\right).
\end{equation}
The resulting formula is given in Ref. \cite{GGGC10},
\begin{equation}
\tilde y = \left(\frac{(\eta/y)^{4/3}+ \kappa^{4/3}I^{1/2}(\eta/y,0)}{\kappa^{2/3}(\eta/y)^{8/3}I^{1/4}(\eta/y,0)}\right).
\end{equation}
It gives the minimum value of $\tilde y$ for which $I(\eta/y,0)>0$ and the small eddies begin to contribute to the turbulent shear stress $\tau_t >0$. In fact for each value of the parameter $\beta_d$ there is a minimum value of $\tilde y$ 
denoted $\tilde y_v$ below which $I=0$.
Only after this minimum does $\tilde y$ increase with $\eta/y$. This gives the end of the viscous layer and the  beginning of the buffer layer and a value of the MVP, $u_v$ at 
$\tilde y_v$. It also gives the value of the fluctuation $w$ at $\tilde y_v$ and we can integrate the differential equations 
for $u$ and $w$, with respect to $y$, to get the form of both functions in the buffer layer, inertial layer and the wake.  Along with the formulas in the viscous layer this gives the full functional form. The differential equations use the spectral information through the full functional form of $I$ and the two parameters $\beta_d$ and $\beta_e$ must be fitted to experimental data. 

\begin{figure*}[t]
\centering
 \includegraphics[width=0.8\textwidth]{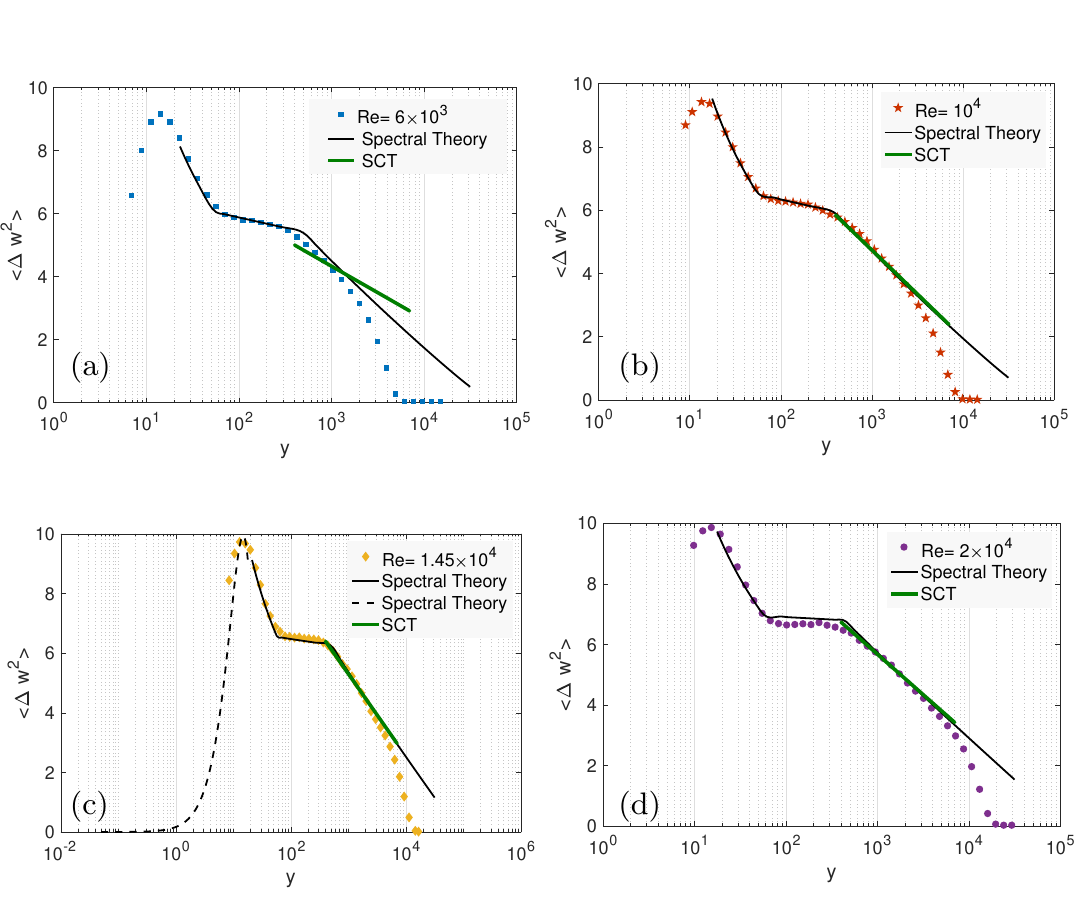}
 \caption{The average of the fluctuation squared as a function of $log(y)$, where $y$ is the distance from the wall (dimensionless units). Comparison of experimental data with theory (blue line).}
  \label{fig:mean variation}
\end{figure*}

Approximations to the MVP and mean square fluctuations, based on the formulas in Sec. \ref{sec:functional} are given in Fig. \ref{fig:mean velocity} and \ref{fig:mean variation}, respectively. To compare with experimental data one must solve the differential equations
\begin{equation}
\label{eq:udiff}
u'=-\frac{1}{2 \kappa^2 I^{3/4}y^2} + \frac{1}{\kappa I^{3/8}y} \sqrt{1 - \frac{y}{Re\sqrt{f}}+\frac{1}{4\kappa^2 I^{3/4}y^2}}
\end{equation}
with the initial condition $u = 4.17$ at the beginning of the buffer layer $y =4.17$. For the fluctuation we first have to solve the differential equation, ignoring term of order $O(1/y^3)$ and higher,
\begin{equation}
\label{eq:wdiff}
w'=\frac{ \sqrt{\tau_0}-\sqrt{\langle\tau_0\rangle}}{ \kappa I^{3/8}y\sqrt{\langle \tau_0 \rangle}} \sqrt{1 - \frac{y}{Re\sqrt{f}}},
\end{equation}
with the initial condition $w=\frac{ \tau_0-\langle\tau_0\rangle}{ \langle \tau_0 \rangle}\left(4.17-\frac{17.39}{2 Re\sqrt{f}}\right)$, from Eq. (\ref{eq:w_o}), at the beginning of the buffer layer. Here $I(y)$ is the integral in Eq. (\ref{eq:spectcont}). 

In practice it is easier to vary the initial conditions than to change $\beta_d$ and $\beta_e$, thus we will let the initial condition $y_o$, of $w$, from Equation (\ref{eq:w_o}), vary slightly depending on the Reynolds number in the simulations below. The other initial condition $w_o$ is given by the formula $w_o=\frac{ \tau_0-\langle\tau_0\rangle}{ \langle \tau_0 \rangle}\left(y_o-\frac{y_o^2}{2 Re\sqrt{f}}\right)$.

\section{Comparison with Experimental Data}
\label{sec:data}

The data we use to compare with the theory comes from the wind tunnel experiments at the University of Melbourne using the nano-scale thermal anemometry probe (NSTAP) to conduct velocity measurements in the high Re number boundary layer up to $Re_\tau  = 20000$. The NSTAT has a sensing length almost one order of magnitude smaller than conventional hot-wire, hence allows for a fully resolved NSTAT measurement of velocity fluctuations, \cite{Sa18}, \cite{Ba19}. The size of the University of Melbourne wind tunnel and the accuracy of the NSTAT permit the measurement over a very large range of scales.  We use the averaged velocity time-series at Reynolds numbers $Re_\tau=6000, 10000,14500, 20000$ and the averaged variance at the same Reynolds numbers. Fig. \ref{fig:mean velocity} shows the mean velocity profiles as a function of normalized distance from the wall, whereas Fig. \ref{fig:mean variation} shows the averaged fluctuation squared (variation) as a function of the normalized distance to the wall. 
Both are semi-log plots. 

First, let us consider the curve describing the MVP in Fig. \ref{fig:mean velocity} (panel b). It starts with the Eq. (\ref{eq:u_o}) for the viscous profile because the $I$-function is zero. But then we reach the value $y_v$ where
the first attached eddies appear ($y=4.17$) and then the viscous profile changes, instead of reaching its maximum $u=Re \sqrt{f}/2$ at $y=Re \sqrt{f}$, the attached eddies increase the viscosity (decrease the Reynolds number) and the MVP reaches its maximum increase at $y \approx 15$, independent of the Reynolds number. The energy transfer of the attached eddies is captured by the $I$-integral and we integrate the differential equation given by Eq. (\ref{eq:udiff}), from $y=4.17$, with the initial condition $u=4.17$. This gives the MVP in Fig. \ref{fig:mean velocity} (b). This was already done in Ref. \cite{GGGC10} and describes how the attached eddies transfer energy 
into the buffer and the inertial layer. However, we notice that in the predicted MVP over estimates the mean velocity in buffer region. This is because the $I$-function from Eq. (\ref{eq:spectcont}) does not account for the formation of the attached eddies which reduce the net energy transfer in the direct cascade. 

The curves for the fluctuations squared in Fig. \ref{fig:mean variation} are obtained in a similar manner. The attached eddies fix the peak of $\langle w^2 \rangle$ at $y \approx 15$ and the peak profiles can be fitted by the viscous formula $\langle w^2 \rangle = a (y-\frac{y^2}{30})^2$ where $a \sim( \langle \tau _o^2\rangle -\langle \tau_o \rangle^2)/ \langle \tau_o \rangle^2$. This fit is shown in Fig. \ref{fig:mean variation} (c). The peak position is experimentally observed to be fixed, but its height shows a weak Reynolds number dependence 
$a = -3.06+0.99 \log(Re)$, see \cite{Sa18}. This relationship can be tested using our theory and this will be done in another publication, see also \cite{CS20}. Then, we integrate the differential equation from Eq. (\ref{eq:wdiff}) for $w$ with the initial data described in last section from some point to the right of the peak, where above peak profile fits the initial condition, this give the profile of the fluctuations squared down to the flat part in the buffer layer. At the beginning of the flat part, $y \approx 60$, the second scaling from Section \ref{sec:MVP} begins to dominate the fluctuations, modeling an inverse cascade of attached eddies in the buffer layer. Then we switch to the buffer $I$-function $I_ b$ in the integration and integrate with $I_b$ until we get into the inertial region where the Kolmogorov-Obukhov scaling dominates again and the attached eddies break up. This produces the curves in Fig. \ref{fig:mean variation}.

We can now compare the functional form of the fluctuations squared shown in Fig. \ref{fig:mean variation}
with the predictions of the stochastic closure theory (SCT) of turbulence, used in Refs. \cite{BC16} and \cite{KBK19}, to compute the Townsend-Perry constants, in the inertial (log) layer. These computations use the first structure function $S_1$ of turbulence and we explain how they are performed, see  \cite{BC16} and \cite{KBK19} for more information. The computed Townsend-Perry constants are listed in Table I. 

The first structure function of turbulence is, see \cite{KBBS17},
\begin{eqnarray*}
	&&E(\vert u(x,t)-u(y,t)\vert)=S_1(x,y,t)\no\\
	 &&=\frac{2}{C}\sum_{k\in\mathbb{Z}^3\backslash\{0\}}\frac{\vert d_k\vert(1-e^{-\lambda_kt})}{\vert k\vert^{\zeta_1}+\frac{4\pi^2\nu}{C}\vert k\vert^{\zeta_1+\frac{4}{3}}}\vert \sin(\pi k\cdot(x-y))\vert,
\end{eqnarray*}
where the Reynolds number dependence enters through the viscosity $\nu$, and $E$ denotes the expectation (ensamble average).
To get the Kolmogorov-Obukhov coefficients, $C_p$ in 
\begin{equation}
S_p(r, \infty) \sim C_p r^{\zeta_p},
\end{equation}
for the lag variable $r$ small, and $\zeta_p$ the scaling exponents, we send $t$ to $\infty$ in the above formulas
and project onto the longitudinal lag variable ${\bf r} = (r,0,0)$. For $p=1$ this becomes
\begin{align}
&&S_1 \sim \frac{2\pi^{\zeta_1}}{C} \sum_{k\neq 0} \frac{|d_k |}{(1+\frac{4\pi^2\nu}{C}|k|^{4/3})} r^{\zeta_1}\no\\
&=&\frac{4\pi^{\zeta_1}}{C} \sum_{k = 1}^\infty \frac{a}{(a^2+k^m)(1+\frac{4\pi^2\nu}{C}|k|^{4/3})} r^{\zeta_1},
\end{align}
see \cite{KBBS17}, where $\zeta_1 = 0.37$, see \cite{BB211}. Now we use the values for $\nu$ in Table 1 in \cite{KBK19}, and the corresponding values 
for $a,\ m$ and $C$ from Table 3 in the same paper. The Reynolds numbers, 6430, 10,770, 15,740 and 19,670 are close enough to ours 6000, 10,000, 14,500, and 20,000, that we can use value of the parameters in \cite{KBK19}. This gives the values in Table I, where $A_1 \sim K|y^*|^{\zeta_1} C_1$, see Section \ref{sec:SCT}, and the proportionality factor 
$K|y^*|^{\zeta_1} = 1/12.952$ is computed at the Reynolds number $15,470$, where the approximated $A_1$ coincides with the measured $A_1$. The $\log$ functions with coefficient $A_1$, from the third column in Table I, and using the constant $B_1$ from the fourth column in Table I, are then compared to the experimental and theoretical values in Fig. \ref{fig:mean variation}. The spanwise Townsend-Perry constants, for the spanwise fluctuations, can computed similarly by projecting onto the spanwise lag variable ${\bf t}=(0,t,0).$

In  Fig. \ref{fig:mean variation} panel (a), the Townsend-Perry constant $A_1$ computed by the SCT does
not agree with the measured slope. This was already observed in Ref. \cite{KBK19}, since for low Reynolds numbers the $C_1$s do not provide a good approximation to the $A_1$s. They only do for large Reynolds numbers and the 
discrepancy (a) occurs at the smallest Reynolds number. This does not happen for the Generalized Townsend-Perry constants, the reasons are explained in Ref. \cite{KBK19}, and for them the $C_p$s, $p \ge 2$, provide good approximations to the $A_p$s for all Reynolds numbers. 

\begin{table}
\begin{center}
\begin{tabular}{ | l | l | l | p{1.0cm} |}
\hline
$Re_\lambda$ & $C_1$ & $A_1$ & $B_1$  \\ \hline
6000&\ 9.449&0.730& 9.373 \\ \hline
10,000&15.628&1.207&13.073  \\ \hline
14,500&15.500&1.197&13.573  \\ \hline
20,000&14.994&1.158&13.673  \\ \hline
\end{tabular}
\caption{Here, the approximate $A_1$ value is computed from $C_1$ using the proportionality factor
$A_1=C_1/(K|y^*|^{\zeta_1})=C_1/12.952$.}   
\end{center}
\end{table}

\begin{figure}[t]
\centering
  \includegraphics[width=.4\textwidth]{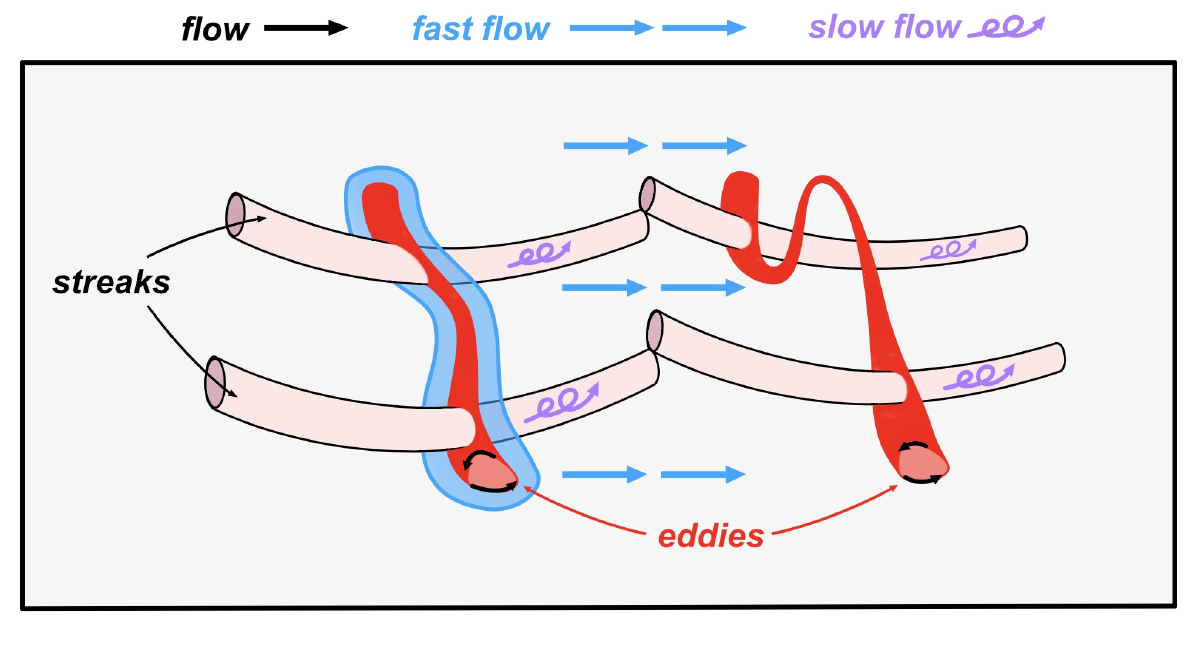}
  \caption{Sketch of the instantaneous streaks, in the streamwise direction, and the wall-attached eddies, in the spanwise direction.}
  \label{fig:wx1}
\end{figure}
\section{Discussion}
\label{sec:summary}
We used the spectral theory of the MVP and the variation profile to represent both, and compare with experiment \cite{Sa18} for a range of Reynolds numbers. Assuming that the wall shear stress is a fluctuating quantity, we can derive that log-law for the variation (\ref{eq:l-lfluct}) that was proposed by Townsend and measured by Perry and Chong. This law involves the Townsend-Perry constants.
This was first done in the large Reynolds number limit and then for general Reynolds numbers. The Reynolds number dependence of the Townsend-Perry constants is determined by the stochastic closure theory \cite{BC16}, \cite{KBK19}. We derive the log-law for the higher moments of the fluctuations and the Generalized Townsend-Perry constants based on the functional form of the variation and use the stochastic closure theory to express them in terms of the Kolmogorov-Obukhov coefficients of the structure functions of turbulence \cite{KBBS17}. This confirms the results in Refs. \cite{BC16} and \cite{KBK19}.

The spectral function $I$ derived in Ref. \cite{GGGC10} plays a central role in this theory. It can be considered be the analytic expression of Townsend's theory of wall-attached eddies. It quantifies when the first eddies appear at the boundary of the viscous and the buffer layer and when they are fully developed in the inertial layer. It even quantifies the limit of their influence in the energetic wake. By introducing the spectral theory into the analysis it resolves many of the issues that we are faced with in boundary layer turbulence. 

The $I$-function corresponds to the Kolmogorov-Obukhov cascade $k_x^{-5/3}$ in the inertial layer, but in the buffer layer another cascade $k_x^{-1}$ dominates the fluctuations, although its influence on the MVP is small. This is an inverse cascade that can accelerate larger and larger attached eddies. The energy transfer of this cascade is captured by the $I$-function in buffer layer, $I_b$. With it we are able to produce the functional form of the averaged fluctuations square in the buffer layer. Once in the inertial layer the original $I$-function dominates again. 

The final confirmation of this spectral theory is how we are able to improve the fit to experimental values of the MVP in Ref. \cite{GGGC10}, by use of the $I_b$ function in the buffer layer. Although,  this effect on the MVP is small, the attached eddies, siphon a small amount of energy from the MVP in the buffer layer. We model this by linear combination of the $I$ and $I_b$ function $(1-a)I+aI_b$, in the buffer layer, where $a$ is small. This produces a better fit to the measured MVP in the buffer region as shown in Figure \ref{fig:mean velocity} (a), whereas the fit without this linear combination, shown in Figure \ref{fig:mean velocity} (b), is not as good. 

It is fair to ask what the Townsend attached eddies actually look like since our spectral method is based on them. 
Unlike the streamwise streaks and associated vortices that have been visualize since the experiments of Kline et al. in the 1960s, see Refs. \cite{Kl67} and \cite{Ji99}, the attached eddies are difficult to visualize, either in experiments or simulations. We provide a sketch in Fig. \ref{fig:wx1}, where streamwise streaks are visualized gradually lifting from the boundary by the flow, and perpendicular to them are spanwise attached eddies being deformed by the alternating slow and fast streamwise flow into a hairpin vortex. This does happen both in experiments and observations,  see Ref. \cite{MM19}. However, these hairpin vortices are made unstable by the striations in the streamwise flow and the typical attached eddies are irregular in shape, with the general feature of being stretched by the flow and attached to the wall. One must interpret their influence in a statistical sense. 

\paragraph*{Acknowledgements:} We are thankful to Ivan Marusic, Milad Samie and Christian E. Willert  for kindly sharing with us the wind turbulence experimental data, and Joe Klewicki for useful conversations. We are grateful to Knut Bauer for proving us with the graphic illustrations. This research was supported in part by the National Science Foundation under Grant No. NSF PHY-1748958 through the Kavli Institute for Theoretical Physics.

\bibliographystyle{plain}
\bibliography{referencef}
       
\end{document}